\documentclass[letterpaper, 10 pt, conference]{ieeeconf}  

\IEEEoverridecommandlockouts                              

\usepackage{graphicx} 
\usepackage{amsmath} 
\usepackage{amssymb}  
\usepackage{array}
\usepackage{quotmark}
\usepackage{balance}

\title{\LARGE \bf
Edge computing in 5G cellular networks for real-time analysis of electrocardiography recorded with wearable textile sensors 
}

\author{Nicolai Spicher$^{1}$, {\em Member, IEEE}, Arne Klingenberg$^{1}$, {\em Student Member, IEEE}, \\ Valentin Purrucker$^{1}$, and Thomas M. Deserno$^{1}$, {\em Senior Member, IEEE} 
\thanks{$^{1}$Peter L. Reichertz Institute for Medical Informatics of TU Braunschweig and Hannover Medical School,
        38106 Braunschweig, Germany. Corresponding author: {\tt\small nicolai.spicher@plri.de}}%
}

\begin{document}

\maketitle
\thispagestyle{empty}
\pagestyle{empty}

\begin{abstract}

Fifth-generation (5G) cellular networks promise higher data rates, lower latency, and large numbers of interconnected devices. Thereby, 5G will provide important steps towards unlocking the full potential of the Internet of Things (IoT). In this work, we propose a lightweight IoT platform for continuous vital sign analysis. Electrocardiography (ECG)  is acquired via textile sensors and continuously sent from a smartphone to an edge device using cellular networks. The edge device applies a state-of-the art deep learning model for providing a binary end-to-end classification if a myocardial infarction is at hand. Using this infrastructure, experiments with four volunteers were conducted. We compare 3rd, 4th-, and 5th-generation cellular networks (release 15) with respect to transmission latency, data corruption, and duration of machine learning inference. The best performance is achieved using 5G showing an average transmission latency of $\mathbf{110}$ms and data corruption in $\mathbf{0.07}$\% of ECG samples. Deep learning inference took approximately $\mathbf{170}$ms. In conclusion, 5G cellular networks in combination with edge devices are a suitable infrastructure for continuous vital sign analysis using deep learning models. Future 5G releases will introduce multi-access edge computing (MEC) as a paradigm for bringing edge devices nearer to mobile clients. This will decrease transmission latency and eventually enable automatic emergency alerting in near real-time.
\end{abstract}

\section{INTRODUCTION}

Cardiovascular diseases are the main cause of deaths worldwide and are responsible for approximately 18 million death each year. In case of an acute myocardial infarction, an immediate response increases probability of survival significantly. However, victims often are unable to call for help and multiple studies reported delays in emergency calls by first-aiders \cite{Takei_2010}. This underlines the need for fully automatic emergency alerts that need to be built upon a reliable infrastructure \cite{Spicher_2021}.

Recently, textile sensors have been proposed for monitoring of vital signs that are woven into stretchy fabrics, allowing unobtrusive and continuous measurements \cite{Khundaqji_2020, Arquilla_2020}.
The availability of large training data \cite{Wagner_2020} lead to the development of deep learning methods, e.g., convolutional neural networks (CNN), showing strong performance in ECG classification \cite{Strodthoff_2020}. However, such complex algorithms require efficient and fast processing, which is usually not possible on mobile devices. 

Proposed by the $3^{\rm rd}$ Generation Partnership Project (3GPP), Relase 15 of the fifth generation (5G) cellular network standard is currently deployed   \cite{Ghosh_2019}. Future revision will provide new technologies, namely enhanced Mobile Broadband (eMBB), ultra-reliable low latency communication (URLLC), and massive machine-type communication (mMTC) \cite{Shafi_2017}. eMBB aims for user experienced data rates reaching $1$ Gbit/s, URLLC for an over-the-air latency as low as $1$ ms, and mMTC for $10^6$ clients per square kilometer.

Moreover, future releases will introduce 5G-powered MEC. This principle substitutes centralized cloud computing by directly processing the data where it is produced: at the edge of the network \cite{Liu_2020}. This is seen as a catalyst for the development of the IoT which embraces all kinds of electrical devices with connectivity that are embedded in smart homes \cite{Wang_2021}, cars \cite{Storck_2019}, or wearables \cite{Baker_2017}. The combination of IoT technology with 5G MEC will significantly reduce transmission latencies and increase security, which will considerably transform healthcare processes \cite{Pham_2020}.

In this work, we build upon these recent developments and propose a platform for end-to-end classification of ECG signals, which are acquired using textile sensors and continuosly transmitted via smartphone to an edge device for real-time analysis.
\section{MATERIAL AND METHODS}

\begin{figure}[h]
    \centering
       \fbox{ \includegraphics[width=.94\linewidth]{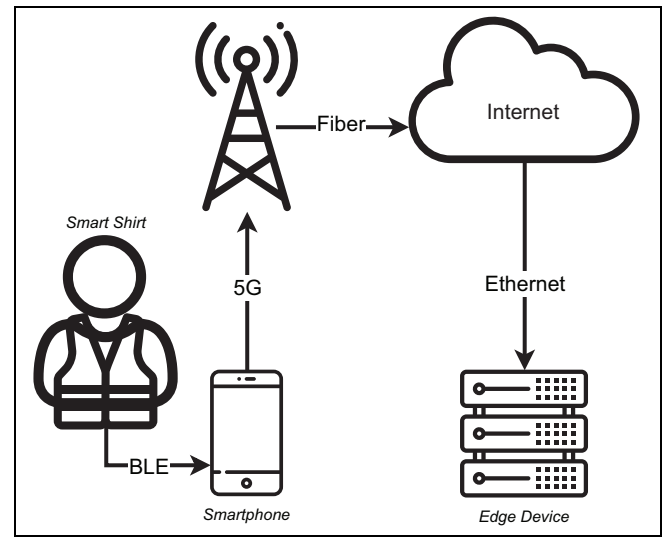} }
            \caption{Proposed architecture: Vital data is sent continuously via Bluetooth Low Energy (BLE) from a smart shirt to a 5G-enabled smartphone which forwards the data to an edge device for end-to-end ECG classification. Icons are freely available from https://www.flaticon.com/. See Acknowledgment.}
            \label{architecture}
\end{figure}

The proposed architecture is composed of a smart shirt, a mobile application, an edge computing device, and a CNN-based algorithm for real-time analytics (Fig. \ref{architecture}).

\clearpage

\subsection{Smart shirt}
Biosignals are acquired using an elastic fabric with integrated textile sensors ({Hexoskin ProShirt}\footnote{https://www.hexoskin.com}, Carré Technologies, Canada). Although not delivered as a medical device, the shirt showed adequate accuracy \cite{vollmer_2019}. It records a single-lead ECG signal ($256$ Hz), thoracic and abdominal respiration ($128$ Hz each), and accelerometry in all three dimensions ($64$ Hz each). In this work, we consider the ECG signal only.

\subsection{Mobile application}
We developed a custom mobile application for the Android operating system ($\ge$ v5.0) incorporating a commercially-available software development kit (SDK) provided by the vendor of the smart shirt. After starting the mobile application and connecting to the smart shirt, data is transmitted continuously in real-time via  BLE using batches of $16$ ECG samples. 
In a parallel process, the application serves as a Message Queuing Telemetry Transport (MQTT) client forwarding the data to the edge device using the Eclipse Paho Android Service library\footnote{https://www.eclipse.org/paho/index.php?page=clients/android/index.php}. 

\subsection{Edge device}
The {Jetson Xavier NX Developer Kit}\footnote{https://developer.nvidia.com/embedded/jetson-xavier-nx-devkit} (6-core NVIDIA Carmel ARM 64-bit CPU, 8 GB RAM, NVIDIA Volta GPU; power mode 15 W; NVIDIA Corporation, CA, USA) serves as edge device featuring the vendor-provided operating system based on Ubuntu Linux. The device is located within the network of a technical university and serves as MQTT broker using Eclipse Mosquitto\footnote{https://www.mosquitto.org/}. We used the out-of-the-box configuration without transmission encryption.

\subsection{Data analytics}

We re-implemented the deep learning neural network architecture proposed by Acharya et al. \cite{Acharya_2017} using Python3 and GPU-enabled Tensorflow\footnote{https://www.tensorflow.org/} and Keras\footnote{https://www.keras.io/}. This CNN architecture with $11$ layers provides a binary decision whether a myocardial infarction is detected in short single-lead ECG signals or not. We adjust the sampling rate ($256$Hz) and process signal length signals of $10$sec only. Training was performed on Google Colab\footnote{https://colab.research.google.com/} before the model was transferred to the edge device.

\section{EXPERIMENTS}

We compare 3G (Universal Mobile Telecommunications System (UMTS)), 4G (Long Term Evolution (LTE)), and 5G cellular networks in their capabilities serving as infrastructure for the proposed architecture with respect to transmission latency, data loss, and inference duration.

\subsection{Experimental design}
We perform three experiments on an empty parking lot in a medium-sized German city. Before each experiment, we synchronized the time on all devices using the network time protocol (NTP). To ensure that no temporal effects bias our results (e.g., load on the cell tower), we perform experiments in parallel (Table \ref{tbl_experiment}). We divide each experiment into two parts: 3G vs.\ 5G and 4G vs.\ 5G. To assess effects of session initiation, each part consists of three runs of $7$ min. In total, we acquired data with a duration of $126 $min.

In each experiment, two subjects in parallel are wearing a smart shirt linked with the mobile application. One application is running on a not-5G-compatible smartphone (\textit{OnePlus 5T}; OnePlus Technology, Guangdong, China) while the other is running on a 5G-compatible smartphone (\textit{Pixel 4A 5G}; Google, CA, USA) with similar specifications. Both smartphones are equipped with the same data plan for business customers with unlimited volume (\textit{Business Mobil XL Plus}, Deutsche Telekom AG, Germany). We establish 3G and 4G connectivity manually using Android operating system features. 

\subsection{Study population}
$N=4$ healthy volunteers  (gender: $1$ female, age: $25.2 \pm 6.2$ years, weight: $64.4 \pm 9.6$ kg, height: $171.8 \pm 10.9$ cm; arithmetic mean $\pm$ standard deviation) took part in the experimental evaluation of our edge computing architecture. Written informed consent was obtained from the subjects regarding storage and analysis of collected data.

\begin{table}[t]
\vspace{0.13cm}
\centering
\setlength\extrarowheight{2pt}
\caption{Design of a single experiment. We conduct this experiment three times, resulting in $126$min of total data.}
\resizebox{\columnwidth}{!}{%
\begin{tabular}{|m{2.3cm}||m{.4cm}|m{.4cm}|m{.4cm}||m{.4cm}|m{.4cm}|m{.4cm}|}
\hline 
 & \multicolumn{3}{c||}{ \bf{Part 1}} &   \multicolumn{3}{c|}{\bf{Part 2}}  \\ \hline\hline
\bf{Smartphone A} & 3G & 3G & 3G & 4G & 4G & 4G \\ \hline
\bf{Smartphone B}& 5G & 5G & 5G & 5G & 5G & 5G \\ \hline
\bf{Duration (min.)}   & ~$7$  & ~$7 $ & ~$7$  & ~$7$  & ~$7$  &  ~$7$    \\ \hline
$\sum$ \bf{Duration (min.)}    & \multicolumn{3}{c||}{ {$21$}} &   \multicolumn{3}{c|}{$21$}  \\ \hline
\end{tabular}
}
\label{tbl_experiment}
\end{table}

\section{RESULTS}

We did not observe any abnormalities (e.g., application crashes) on the smartphones during experiments. On both smartphones, the system load was low. It was ensured that both smartphones were connected with the same cell tower by comparing the cell id.

\begin{figure}[t]
    \centering
\vspace{0.13cm}
       \includegraphics[width=.96\linewidth]{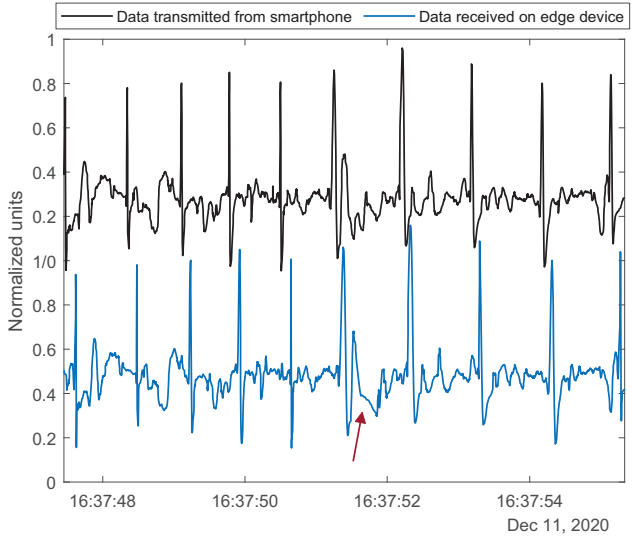} 
            \caption{Example of latency between smartphone and edge device and data corruption. As can be seen, signals received on the edge device are delayed due to transmission. The red arrows indicates corrupted data on the edge device. The range of amplitudes of both signals is identical but the y-axis was split by vertically moving the received signal to increase visibility. }
            \label{signals}
\end{figure}

We perform the evaluation retrospectively by comparing MQTT broker log-files and MQTT client log-files stored on the smartphones (Fig. \ref{signals}). As the data is received from the smart shirt and forwarded to the MQTT broker in batches of $16$ samples, we perform linear interpolation to compute a single timestamp for each ECG sample received on the edge device. Two log-files on the edge devices are excluded from analysis as they were corrupted.

\subsection{Results of transmission latency}
\begin{figure}[t]
    \centering
       \includegraphics[width=.99\linewidth, trim=30 0 35 0]{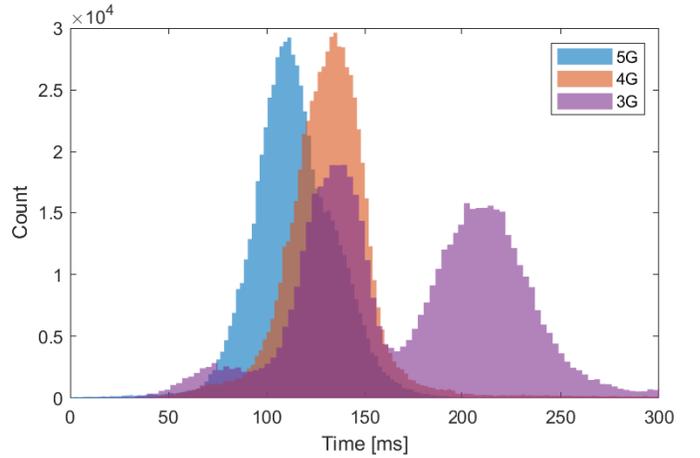} 
            \caption{Histograms of transmission delays from smartphone to edge device using  3G/4G/5G cellular networks. Data shown is averaged over all conducted experiments.}
            \label{histogram}
\end{figure}

By assigning received ECG sample values on the edge device with values sent by the smartphone, we compute their transmission delay (Fig.~\ref{histogram}). The 3G latency distribution is a bimodal distribution with two peaks centered at $137$ms, $210$ms, respectively. The second peak is associated with sudden changes in the transmission delay we observed in the data. The 4G and 5G latency distributions are both approximately Gaussian showing a mean of $134$ms and $114$ms, respectively.

\subsection{Results of data corruption or loss}
\label{sec:data_loss}
Additionally, we compute the number of missed or corrupted ECG samples using a heuristic approach. We align sent and received samples and use a sliding window approach to detect unequal or missing values. The red arrow in Fig. \ref{signals} indicates corrupted ECG samples on the edge device. The average number of missing or unequal ECG samples are $2.98 \pm 6.23$\% (3G),  $0.85 \pm 1.4$\% (4G),  $0.07 \pm 0.06$\% (5G). 

\subsection{Results of inference duration}
\begin{figure}[t]
    \centering
       \includegraphics[width=.99\linewidth, trim=20 0 35 0]{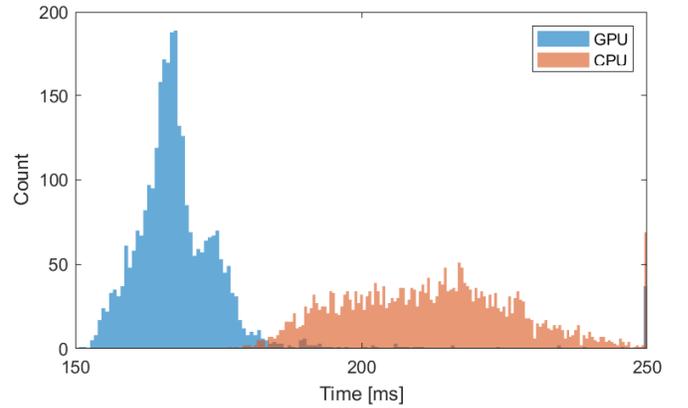} 
            \caption{Histogram of ECG inference durations in segments of $10$sec. Values larger than $250$ms (GPU: $1.2$\%, CPU: $2.1$\%) on the x-axis are clamped to $250$ms to increase visibility. }
            \label{histogram2}
\end{figure}

We apply the deep learning model to the ECG data after the experiments. We feed all received data in two parallel processes to the pre-trained Keras model in segments of $10$sec and store the duration of inferencing (Fig.~\ref{histogram2}). 

Using GPU support, more than $98$\% of values are in the range of $150-180$ms with a peak at approximately $165$ms. We did not observe effects of GPU \tqt{warm-up}. Disabling the GPU and using CPU only, inferencing is almost always slower than the GPU and less stable, resulting in a broad distribution reaching maximum durations up to $250$ms.

\section{DISCUSSION}

\subsection{Results}

Outdoor experiments with four volunteers at a stationary position show that transmission delays of approximately $110$ms and GPU inference delays smaller than $180$ms can be reached. It should be noted that we did not fine-tune the MQTT connection or deep learning architecture. Therefore, reported results can possibly be improved.

However, we cannot estimate the BLE transmission delay between smart shirt and smartphone and therefore the real delay between ECG sensor activation to classification is slightly larger. BLE delays as low as $5$ms have been reported in literature which reach -- in worst-case scenarios -- values up to approx $50$ms \cite{Rond_n_2017}. Therefore, we believe that total duration can be kept below $300$ms using the proposed infrastructure in combination with 5G.

Certainly, our base functionality poses limitations. No security mechanisms, such as encrypted data transmission or user authentification were implemented. We aimed for developing the \tqt{core} platform, which serves as bottom base-line with respect to functionality. In future work, we will add features and analyze their influence on measured parameters. 

Additionally, experiments were performed at a stationary location. Therefore, the influence of switching between different cell towers or dead spots with no reception need to be addressed.  Furthermore, we only processed two single-lead ECG signals in parallel on the edge device. Hence, the impact of higher data load on performance needs to be evaluated. 

Regarding ECG analysis, we used a pre-defined deep learning model for detection of myocardial infarction \cite{Acharya_2017}. Training data was acquired from freely-available databases measured with conventional ECG devices \cite{Goldberger_2000}. However, it is not guaranteed that the ECG signal measured via a textile sensors has the same morphology. Furthermore, motion artefacts pose a serious problem that might lead to false classification. Therefore, our future research aims at analyzing these aspects as well as adding other ECG analysis methods, e.g., delineation  enabling the measurement of clinically relevant intervals \cite{Spicher_2020}.

\subsection{Limitations}
Our data analysis has certain limitations with respect to accuracy. Aligning time-delayed data from multiple sensors with potential data loss or corruption is a non-trivial task \cite{vollmer_2019}. Although we confirmed our results manually, our analysis may be biased. However, as signals from all experiments were processed by the same algorithm, the order of decreasing data corruption/latency from 3G over 4G to 5G should be maintained even if the heuristic is biased.

Additionally, two different smartphones were used. However, both have similar specifications and the developed app has only minimal hardware requirements. Therefore, we do not expect a significant bias due to the different hardware.

Furthermore, it should be added that a fundamental issue of the proposed archiecture is the susceptibility to the cellular network coverage and energy consumption due to data transfer. Deploying the ECG analysis on the smartphone by means of a finely-adjusted CNN would be a more reliable solution w.r.t. these aspects. However, there are also cons like increased energy consumption due to inference. Such an approach could serve as a valuable "fallback" method in case of celullar dead zones.

\subsection{Outlook}
In this work, 5G networks in current 3GPP Release 15 building upon existing 4G infrastructure (\tqt{Non-stand alone mode}) were used. Future releases will enable the \tqt{Stand alone mode} and introduce URLLC and eMBB, enabling even lower latencies and higher data rates, respectively. 

\begin{figure}[t]
    \centering
\vspace{0.13cm}
       \fbox{ \includegraphics[width=.71\linewidth]{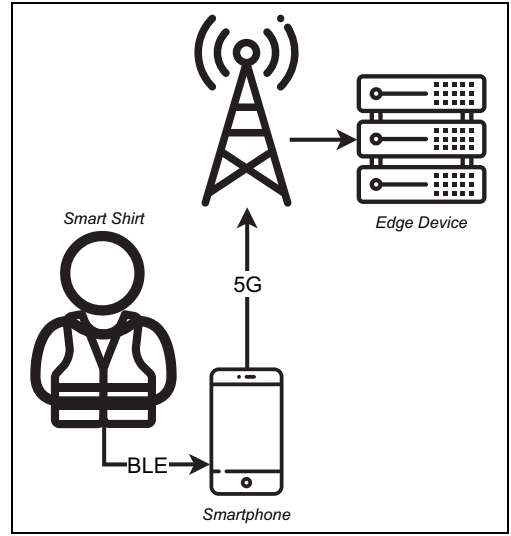} }
            \caption{Envisioned architecture: Powered by the principles of 5G MEC the edge device can be brought into proximity of the smartphone, thereby decreasing latency. Furthermore, data is not sent via the unsafe channel internet, increasing data security. The arrow between cell and edge device is not labeled as there is not de-facto MEC standard, yet.}
            \label{architecture2}
\end{figure}

For the deployment of the edge device we used a conventional \tqt{cloud} architecture over the internet. New technologies such as 5G-enabled MEC \cite{Taleb_2017} or network slicing \cite{Kapassa_2019} will introduce new features with potential value for the proposed platform (Fig.~\ref{architecture2}). Bringing the edge device closer to the smartphone reduces latency and eventually advances fully automatic emergency alerts in near real-time \cite{Spicher_2021}.

\section{CONCLUSION}
The sudden onset of cardiac diseases such as myocardial infarction require an immediate response. We report on an IoT platform which enables the continuous processing of single-lead ECG signals. Our results show that a delay $\le300$ms from ECG sensor measurement to end-to-end classification can be reached. Further revisions of 5G cellular networks could significantly enhance the proposed architecture.

\addtolength{\textheight}{-12cm}   




\section*{ACKNOWLEDGMENT}
All procedures performed were in accordance with the 1964 Helsinki declaration, as revised in 2000.
This work received funding by the German Federal Ministry of Transport and Digital Infrastructure (BMVI) under grant \#VB5GFWOTUB.
Icons of Fig. \ref{architecture} and Fig. \ref{architecture2} made by Freepik, Kiranshastry, iconixar, srip (https://www.flaticon.com/)

\balance

\bibliographystyle{IEEEtran}

\end{document}